\def\kms{\ifmmode{\,\hbox{km}\,s^{-1}}\else {\rm\,km\,s$^{-1}$}\fi}

\def\msun{{\rm\,M_\odot}}
\def\lsun{{\rm\,L_\odot}}

\def\kmsm{{\rm\,km\,s^{-1}\,Mpc^{-1}}}
\def\kmps{{\rm\,km\,s^{-1}}}
\def\hmpc{\ifmmode{h^{-1}\,\hbox{Mpc}}\else{$h^{-1}$\thinspace Mpc}\fi}

\def\et{{\it et~al.}~}

\def\ie{{\it i.e.}~}
\def\sigp{\ifmmode{\sigma_p}\else {$\sigma_p$}\fi}
\def\sig1{\ifmmode{\sigma_1}\else {$\sigma_1$}\fi}
\def\r200{\ifmmode{r_{200}}\else {$r_{200}$}\fi}
\def\spose#1{\hbox to 0pt{#1\hss}}
\def\lta{\mathrel{\spose{\lower 3pt\hbox{$\mathchar"218$}}
     \raise 2.0pt\hbox{$\mathchar"13C$}}}
\def\gta{\mathrel{\spose{\lower 3pt\hbox{$\mathchar"218$}}
     \raise 2.0pt\hbox{$\mathchar"13E$}}}
\def\bull{\par\noindent\parskip=0pt\hangindent=3pc\hangafter=0 $\bullet$~}

\def\mlobs{289\pm50}

\def\mlclose{1136\pm138}

\def\omzeroc{0.19\pm0.06}

\def\mvadj{0.82\pm0.14}
\def\ladj{0.11\pm0.05} 

\def\apj{ApJ}
\def\apjl{ApJ(Lett)}
\def\physrep{Physics Reports}
\def\apjs{ApJS}
\def\mnras{MNRAS}
\def\aap{AAp}
\def\aj{AJ}

\def\araa{ARAA}

\input{epsf}
\input{rotate}
\documentstyle[editedvolume]{crckapb}

\begin{opening}
\title{The CNOC Cluster Survey}

\subtitle{$\Omega$, $\sigma_8$, $\phi(L,z)$ Results, 
	and Prospects for $\Lambda$ Measurement}

\author{R. G. CARLBERG}
\author{H. K. C. YEE}
\author{H. LIN}
\author{C. W. SHEPHERD}
\author{P. GRAVEL}
\institute{University of Toronto, Toronto ON, M5S~3H8 Canada}
\author{E. ELLINGSON}
\institute{University of Colorado, CO 80309 USA}
\author{S. L. MORRIS}
\author{D. SCHADE}
\author{J. E. HESSER}
\author{J. B. HUTCHINGS}
\author{J. B. OKE}
\institute{National Research Council of Canada,
	Herzberg Institute of Astrophysics,
	Dominion Astrophysical Observatory, 
        Victoria, BC, V8X~4M6, Canada}
\author{R. ABRAHAM}
\institute{Institute of Astronomy, Cambridge CB3~OHA, UK}
\author{M. BALOGH}
\author{G. WIRTH}
\author{F. D. A. HARTWICK}
\author{C. J. PRITCHET}
\institute{University of Victoria,
        Victoria, BC, V8W~3P6, Canada}
\author{T. SMECKER-HANE}
\institute{University of California, Irvine,
        CA 92717, USA \\}

\end{opening}

\runningtitle{THE CNOC CLUSTER SURVEY}

\begin{document}

\begin{abstract}
Rich galaxy clusters are powerful probes of both cosmological
and galaxy evolution parameters. The CNOC cluster survey
was primarily designed to distinguish between $\Omega=1$ and
$\Omega\simeq0.2$ cosmologies.  Projected foreground and background
galaxies provide a field sample of comparable size. The results
strongly support a low-density universe.  The luminous cluster
galaxies are about 10-30\% fainter, depending on color, than the
comparable field galaxies, but otherwise they show a slow and nearly
parallel evolution. On the average, there is no excess star formation
when galaxies fall into clusters. These data provide the basis for a
simple $\Lambda$ measurement using the survey's clusters and the field
data. The errors in $\Omega_M$, $\Omega_\Lambda$, $\sigma_8$ and
galaxy evolution parameters could be reduced to a few percent with a
sample of a few hundred clusters spread over the $0<z<1$ range.
\end{abstract}

\section{The $\Omega$ Problem}

The total mass-to-light ratio of field galaxies multiplied with the
field luminosity density gives, by definition, the mean mass density
of the universe, $\rho_0$ \cite{oort,gunn}.  The complications in the
measurement are to provide unbiased mass estimates and to measure any
luminosity difference between the galaxies for which total system
masses are known and field galaxies. Small, dense systems, like
individual galaxies and small groups, exhibit ``luminosity
segregation'' where the luminosity distribution is more concentrated
than the mass distribution \cite{bld}. The largest virialized
structures --- rich clusters --- indicate $\Omega\simeq 0.2-0.4$, but
these measurements may be biased by luminosity segregation and
differential evolution of the very red cluster galaxies as compared to
bluer field galaxies.  On even larger scales, where overdense
structures are still expanding, but retarded from the Hubble flow,
bulk flows of galaxies measure the parameter $\Omega^{0.6}/b$, where
$b$ is the assumed linear bias, $b=(\delta n/n)/(\delta\rho/\rho)$,
relating the tracer galaxy number densities and the mass field
density.  The flow measurements tentatively indicated $\Omega\simeq1$
\cite{dekel,sw}, but recent increases in sample size and new analyses
indicate possible compatibility with $\Omega\simeq 0.2-0.4$
\cite{iras,dnw,wsdk}. The value $\Omega=1$ is of considerable interest
as the ``natural'' value (the ``Dicke coincidence'') for a universe
that has expanded many e-folds, and as the value originally predicted
by inflationary cosmology \cite{guth,bst,turner}.  If $\Omega=1$ and
all the dark matter falls into clusters along with galaxies (as it
must if it is collisionless and cold), then rich clusters, which draw
their mass from regions 10\hmpc\ in radius, must have a total
virialized mass-to-light ratio, $M_v/L$, about 2.5 to 5 times higher
than the standard virial analysis gives. The cluster data did not rule
out such high $M/L$ values in the field, due to unconstrained
systematic errors in both $M_v$ and $L$.

The Canadian Network for Observational Cosmology (CNOC) sample was
designed to create a dataset that allows complete internal control of
most aspects of the cluster $\Omega$ estimate, especially luminosity
segregation and differential evolution of cluster galaxies relative to
the field.  The cluster sample was chosen from the Einstein Medium
Sensitivity Survey Catalogue of X-ray clusters \cite{emss1,emss2,gl}
to have a high X-ray luminosity, which helps guarantee that the
clusters are reasonably virialized, suggests that they will contain
many galaxies, and allows other cosmological measurements. The sample
was augmented with A2390 to fill an RA gap.  The clusters are at
$z\sim 0.3$ so they have a significant redshift interval over which
the surrounding field galaxies are nearly uniformly sampled in
redshift.  Observations were made at CFHT in 24 assigned nights in
1993 and 1994, of which 22 were usefully clear.  The primary
catalogues contain Gunn $r$ magnitudes and $g-r$ colors for 25,000
objects, plus radial velocity measurements at an average accuracy of
100~\kms \cite{yec} for a subsample of 2600.  All results are reported
for $H_0=100\kmsm$ and $\Omega_0=0.2$, $\Omega_\Lambda=0$.

The overall CNOC program is encapsulated as follows:
\bull{select clusters from an unbiased X-ray catalogue of large $z$ range,}
\bull{observe clusters out to a radius with
	 overdensity of $200\rho_c$ or lower,}
\bull{measure virial mass-to-light ratio,}
\bull{test and correct virial mass,}
\bull{test for $z$ dependence of $M_v/L$, measure $\phi(L,z)$ for clusters,}
\bull{measure $\phi(L,z)$ for the field sample,}
\bull{measure $n(>M,z)$, the cosmological density of clusters, and,}
\bull{measure the evolution of field clustering and clustering velocities.}

\noindent
The main results from this ongoing program of investigation are
summarized below but discussed in detail in a series of papers. The
observational methods, which are designed to efficiently measure
redshifts of galaxies in the primary sample, are discussed in Yee,
Ellingson \& Carlberg (1996). The measurement of global quantities,
such as the velocity dispersion and virial mass-to-light ratio, is
done in Carlberg, \et\ (1996). The average mass and light profiles are
compared in Carlberg, Yee \& Ellingson (1997) to measure the biases of
the virial mass-to-light ratio and correct the luminosities of cluster
galaxies to field galaxy values.  Carlberg, \et\ (1997a) shows that
the same mass profile can be recovered from two independent and very
different subsamples of the cluster data. The average mass profile is
compared to a theoretical prediction in Carlberg, \et\ (1997b),
finding an impressive agreement. The sample was designed such that the
amplitude of the primordial density fluctuation spectrum could be
measured on cluster scales, with the results given in Carlberg, \et\
(1997b). Some preliminary results of the CNOC field survey, now
comprising about 2/3 of the 5000 primary sample redshifts, are given
below.

\section{CNOC Cluster Masses}

The virial mass is $M_v = 3 G^{-1} \sigma_1^2 r_v,$ where \sig1\ is
the line-of-sight velocity dispersion and $r_v$ is the virial radius
of the observed galaxies \cite{bt}. Deciding which galaxies in
redshift space are cluster members is fundamentally ambiguous.  That
is, a cluster galaxy with a completely reasonable line-of-sight
velocity of $1000~\kmps$ appears in redshift space at 10\hmpc\ from
the cluster's center of mass, far outside the virialized cluster and
intermingled with field galaxies.  This complication increases in
severity as the cluster density declines with projected distance from
the cluster center where our sample is specifically intended to
probe. Our straightforward solution to this problem is to subtract the
mean density of field galaxies in redshift space from the redshift
space of the cluster \cite{global}.  The resulting velocity
dispersions are about 13\% lower than precisely the same cluster data
give without background subtraction.  Gratifyingly, our velocity
dispersions are in excellent agreement with those inferred from the
mean X-ray temperatures \cite{mush}.  Our estimate of $r_v$ uses a
``ringwise'' estimate, rather than the traditional ``pointwise''
estimate \cite{bt}. This allows us to include a selection function for
the roughly rectangular window on the sky which outlines the fields
and it helps to reduce noise in the estimated virial radius. The
ringwise estimator overestimates the virial radius of the clusters
because it does not account for their flattening.  We use the mean of
the ratio of the ringwise to the pointwise estimator, which is about
1.28, to make this correction.  The ringwise estimator includes a
small ``softening'', of one arcsec, to avoid a divergence when two
galaxies are at the same radius.

\medskip
\hbox{
\epsfysize 5.0truecm
\epsffile{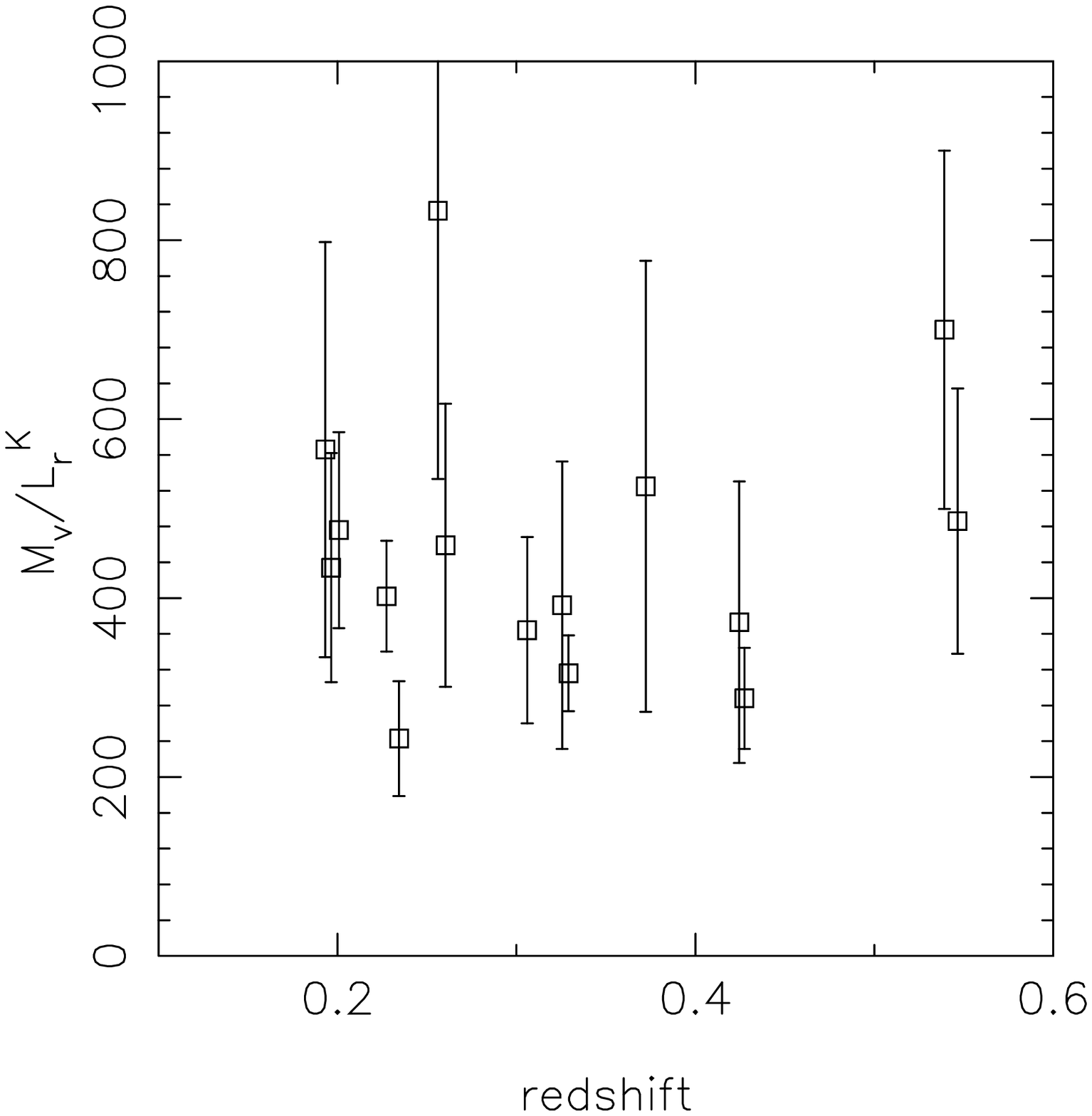}
\quad \vbox to 5truecm{\hsize=7truecm
\vfill\noindent
Figure 1: The virial mass-to-light ratios as a function of
redshift. The luminosities have been k-corrected and allow for
evolution at the rate of $M_\ast(z)=M_\ast(0)-Qz$, with $Q=1$. The 15
clusters are consistent with having a universal $M_v/L(0)=$
$374h\msun/\lsun$ with the same dataset giving a closure value
$1502h\msun/\lsun$, both with 10\% standard errors.
\vfill}}

The total galaxy luminosity of the cluster galaxies in ratio to $M_v$
gives $M_v/L$, an estimate of the total mass-to-light ratio. The
average k-corrected Gunn $r$ $M_v/L$, for $M_r^k\le -19$ mag but
integrated to infinity with the luminosity function ($M_\ast=-20.3$
mag, $\alpha=1.1$), is $\mlobs h\msun/\lsun$ at $z=0.31$.  If
corrected for pure luminosity evolution at the rate of 1 magnitude of
brightening per unit redshift the mean Gunn $r$ $M_v/L$ extrapolates
to $374\pm54 h\msun/\lsun$ at redshift zero.  The mean standard error
per cluster on average is 40\%, although this can be reduced to 25\%
by eliminating the clusters with large errors.  A major result is that
$M_v/L$ is the same for all clusters once minimal passive evolution of
the cluster galaxies is taken into account \cite{global}. For the 15
clusters displayed in Figure~1 $\chi^2=16.6$ which is about 28\%
probable for $\nu=14$ degrees of freedom, that is, consistent with no
intrinsic variation. X-ray \cite{david} and weak lensing analyses
\cite{tf,squires,smail,ft} generally find results in accord with these.

\section{Mass Profiles and Virial Mass Bias}

Our observations extend over what is expected to be the entire
virialized volume of the cluster. Analytic models \cite{gg} and
simulations \cite{cl} find that the virialized mass is generally
contained inside the surface where the mean interior density is
approximately $200\rho_c$, which defines the radius \r200.  The small
extrapolation from the observationally derived $r_v$ to \r200\ assumes
$M(r)\propto r$, which gives $\r200 =
\sqrt{3}\sigma_1/[10 H(z)],$ where $H(z)$ is the Hubble constant at
redshift $z$.

\medskip
\hbox{
\epsfysize 5.0truecm
\epsffile{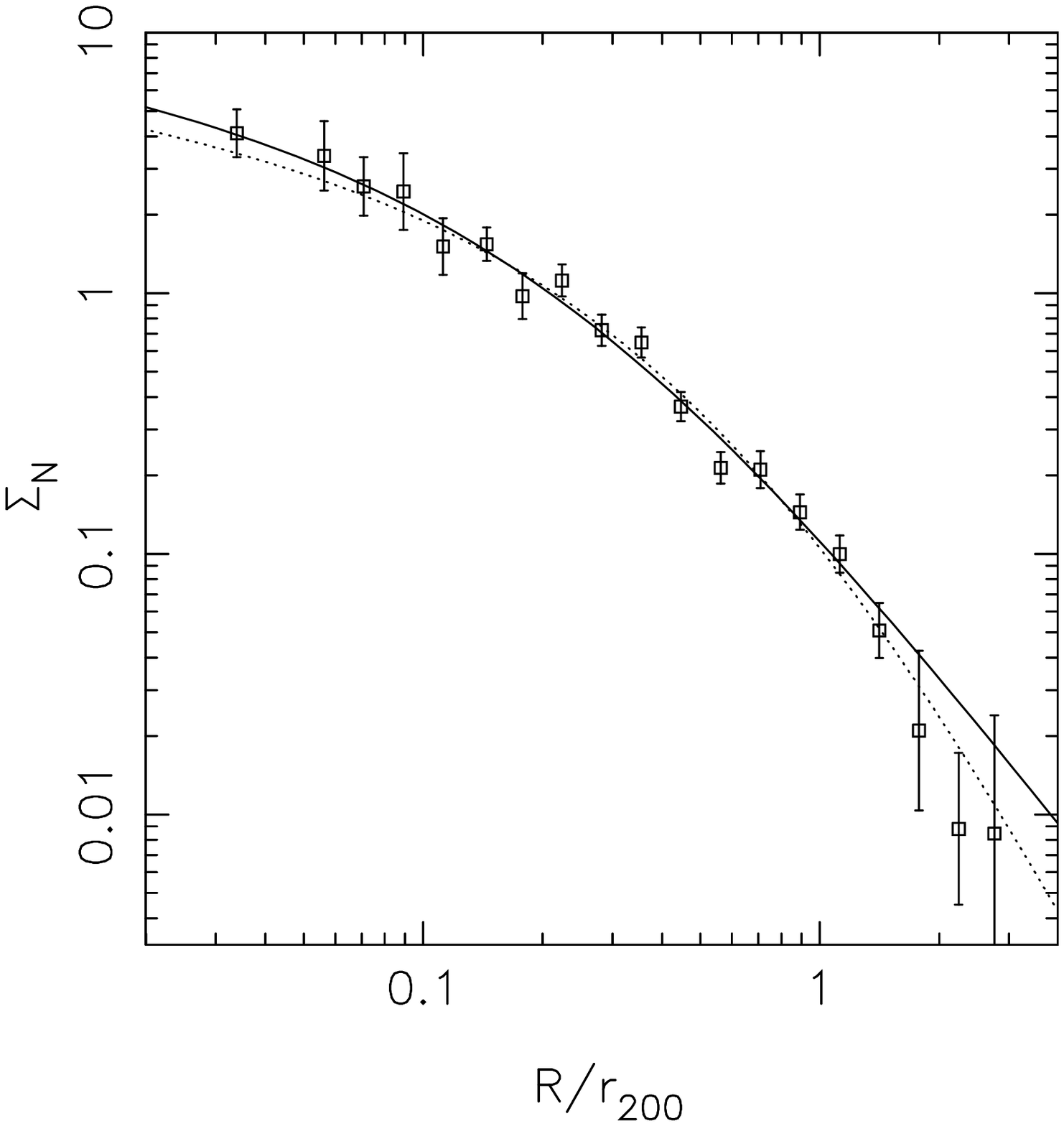}
\quad \vbox to 5truecm{\hsize=7truecm
\vfill\noindent
Figure 2: The background subtracted surface density profile, fitted
with NFW (solid line) and Hernquist (dotted line) functions, both of
which are statistically acceptable. The mass is accurately traced with
this profile, as shown below. The remarkable result is that the NFW
prediction of the scale radius is completely in accord with our
measurements.\vfill}  }

The accuracy of the virial mass depends on the tracer galaxies having
the same distribution as the mass. We compare the light-traces-mass
profile, $M_L(r)=L(r)\times M_v/L$, to the dynamical mass profile,
$M_D(r)$, derived from the Jeans Equation,
\begin{equation}
{\sigma_r^2\over r}
        \left[{{d \ln{\sigma_r^2}}\over{d\ln{r}}} +
        {{d\ln{\nu}}\over{{d\ln{r}}}} +2\beta\right]
        = -{{GM_D(r)}\over{r^2}},
\label{eq:jeans}
\end{equation}
where $\beta = 1 -\sigma_\theta^2/\sigma_r^2$ is the velocity
anisotropy parameter.  This equation does not
depend on the density of the tracer galaxies, $\nu(r)$, following
the mass density profile, $\rho(r)$.

To create an effectively spherical cluster and to reduce the effects
of substructure, all the cluster galaxies are combined into one
``ensemble'' cluster by normalizing the velocities to \sig1\ and the
projected radii to the \r200\ value of each cluster. The two
quantities measured from this distribution are the surface number
density profile of cluster galaxies, $\Sigma_N(R)$, shown in Figure~2,
and the projected velocity dispersion profile, Figure~3. The volume
number density profile is modeled as $\nu(r) = A/[r(r+a)^p], $ where
$p=2$ for the Navarro, Frenk \& White function (1997, hereafter NFW) 
and $p=3$ for Hernquist's (1990) form.

\medskip
\hbox{
\vbox{
\epsfysize 5.0truecm
\epsffile{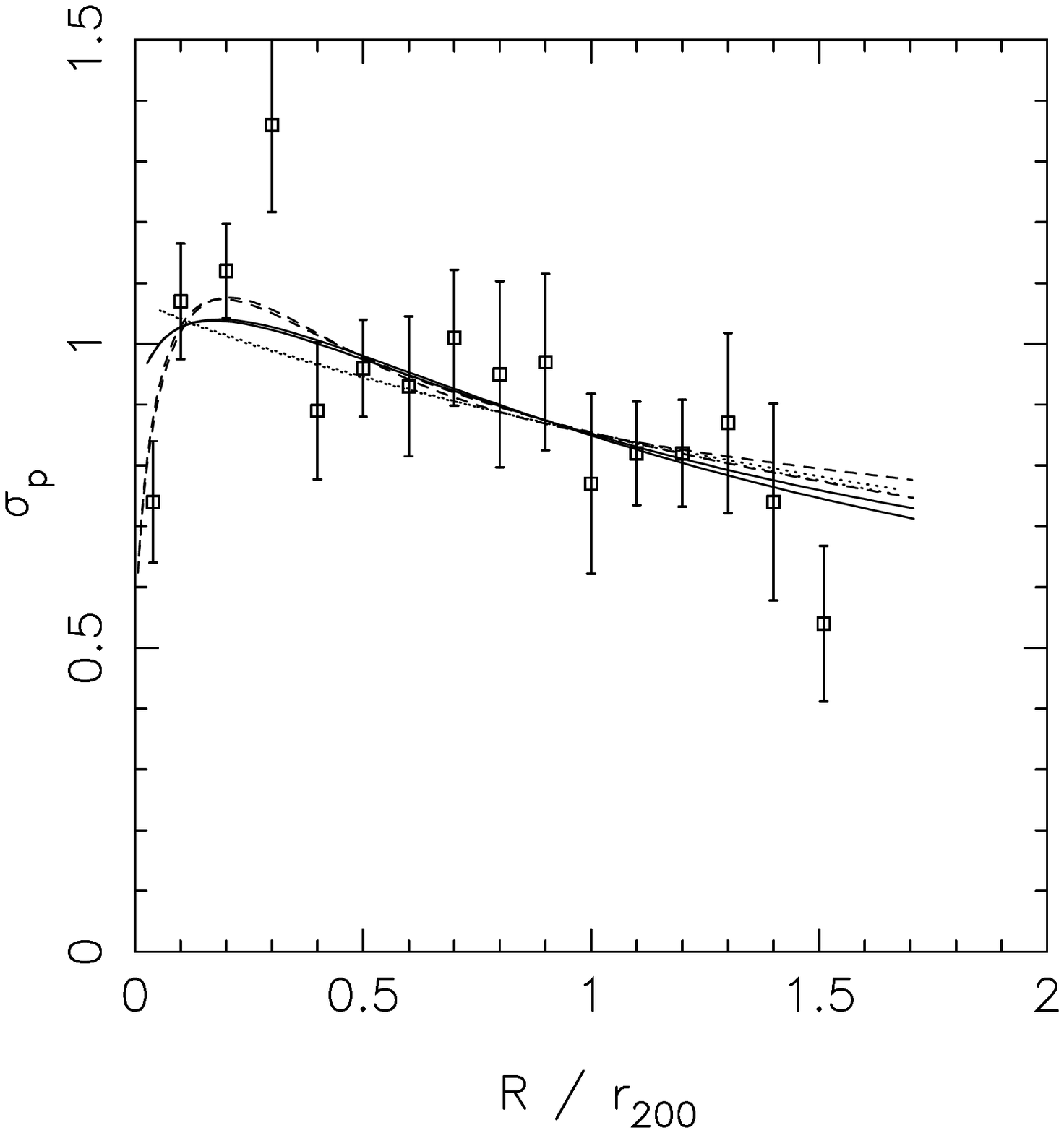}
}
\quad \vbox to 5truecm{\hsize=7truecm\vfill\noindent
Figure 3: The projected velocity dispersion profile and
the projection of the fitted profiles, for a range
of $\beta$ and $[c_1,c_2]$.
See Carlberg \et\ 1997d for details.\vfill
}}

The radial velocity dispersion is modeled as $\sigma_r^2 = B
[c_1r/(1+c_1r) + c_2]/[ 1 + r/b]$, where $B$ and $b$ are the two
parameters adjusted to fit the observed $\sigp(R)$. The $c_1$ and
$c_2$ parameters are externally fixed to vary the shape of the curve.
The velocity anisotropy is taken as an approximate fit to the results
of n-body simulations.  A range of $\sigma_r(r)$ fits are shown in
Figure~3. The resulting ratio, $b_{Mv}(r)=M_D(r)/M_L(r)$ shown in
Figure~4, is not very sensitive to the details of the velocity
modeling.

\medskip
\hbox{
\epsfysize 5.0truecm
\epsffile{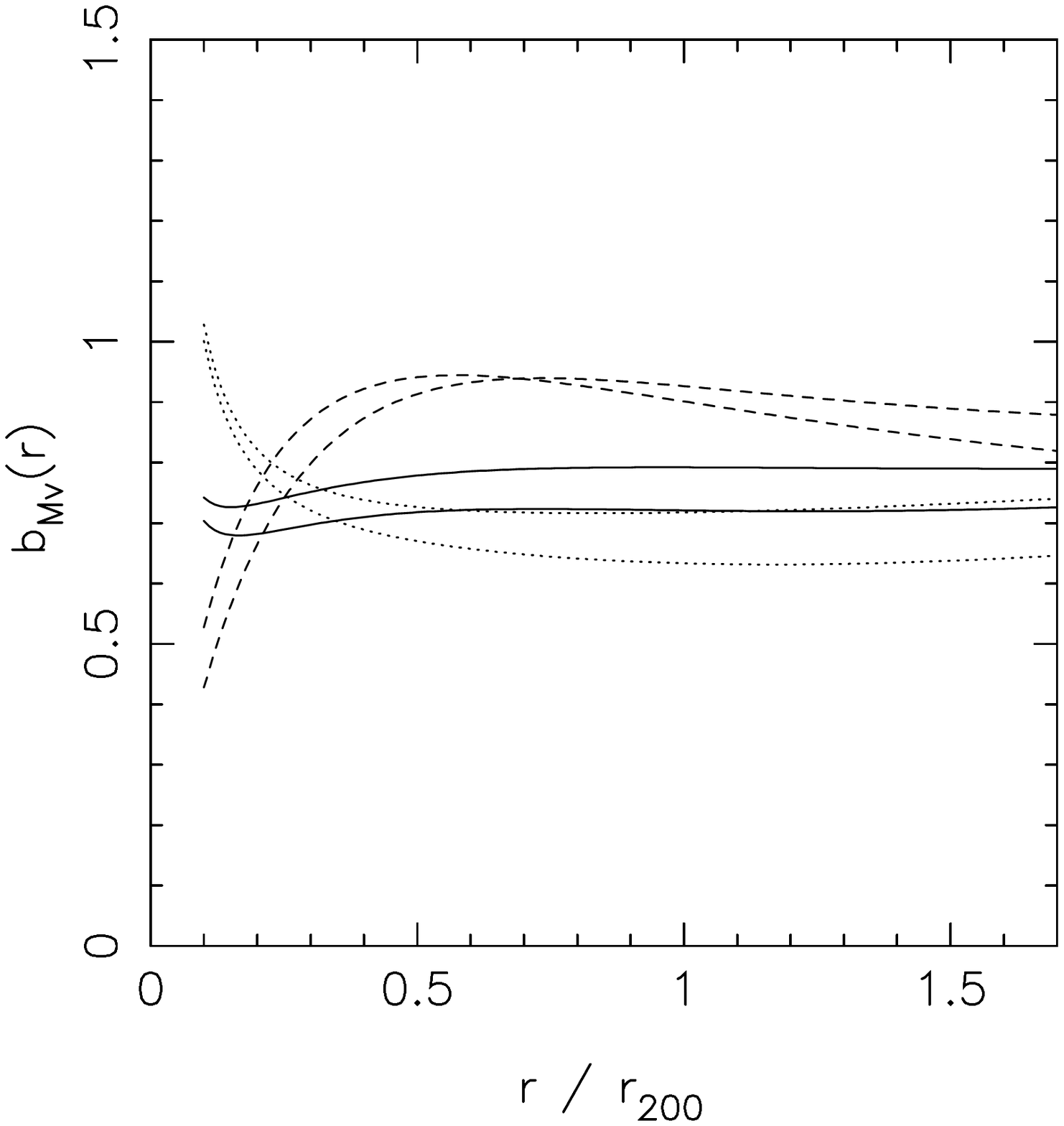}
\quad \vbox to 5truecm{\hsize=7truecm
\vfill\noindent
Figure 4: The derived ratio of the dynamical mass profile, $M_D(r)$,
to $r$ selected galaxy profile, $L(r)$, normalized with the virial
mass-to-light ratio evaluated inside $1.5\r200$.  In each pair of
curves the upper line at small radius is for $\beta_m=0.3$ and the
lower for $\beta_m=0.5$. The dotted line is for $c_1=0$, the dashed
for $c_2=0$ and the solid line is our preferred $c_1=8$, $c_2=1/2$.\vfill  }}

There are two conclusions to be drawn from Figure~4. First, over a
range of about two decades of projected density or three decades of
volume density, the integrated galaxy distribution traces the mass
distribution to an accuracy of about 20\% or better.  Second, the
virial mass is always an overestimate of the true mass contained
within that radius, which we estimate to be a factor $\mvadj$. We
attribute the mass overestimate of $M_v$ to dropping the surface
pressure term at $P_s=0$ in the virial theorem, \ie\ $2T+W=3P_sV$.
Simple modeling of an equilibrium cluster shows that truncating the
data at \r200\ will cause $M_v$ to be upward biased at the 15-25\%
level that is inferred from the Jeans Equation.

\medskip
\vbox{
\hbox{
\epsfysize 4.0truecm
\epsffile{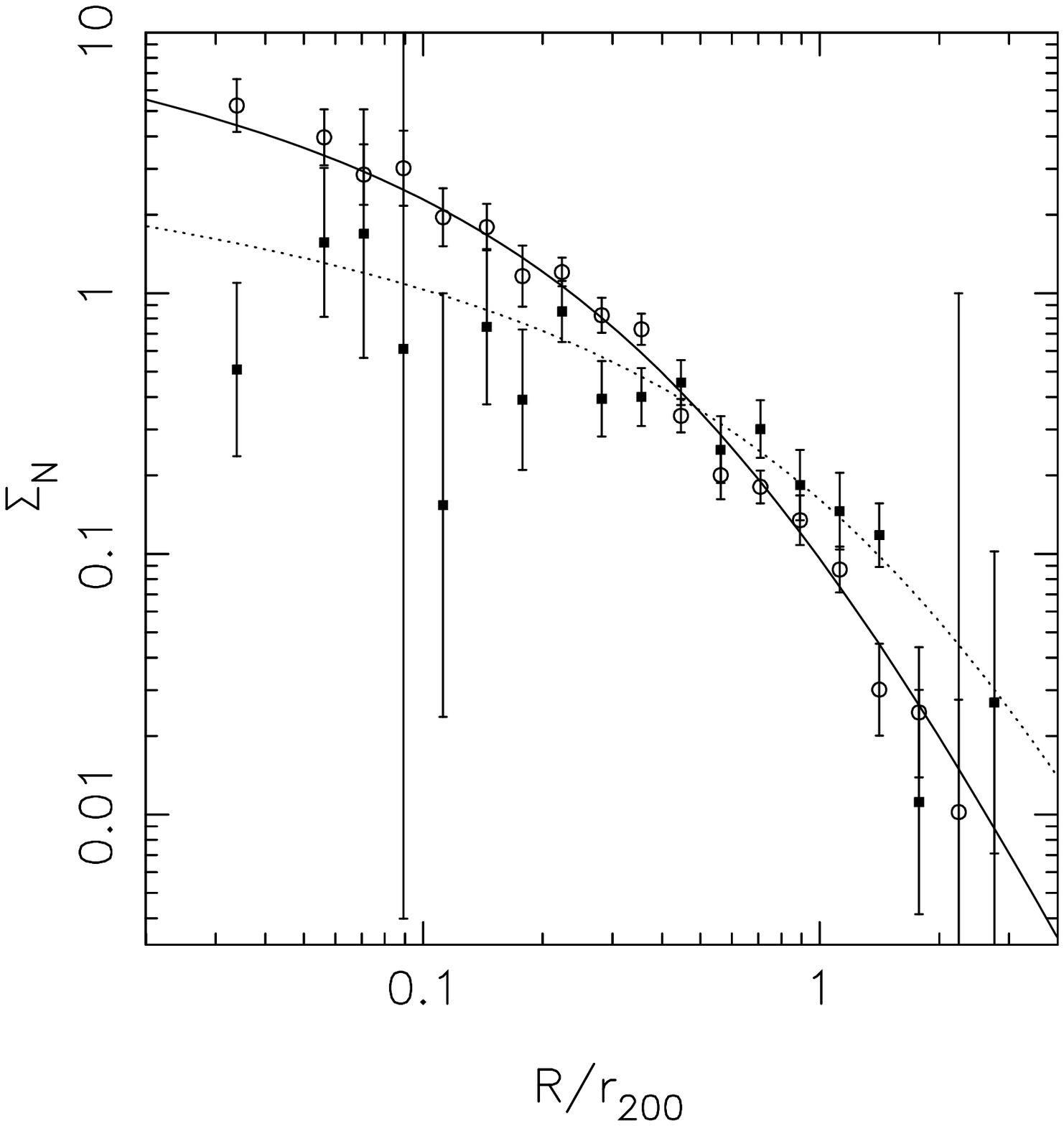}
~~
\epsfysize 4.0truecm
\epsffile{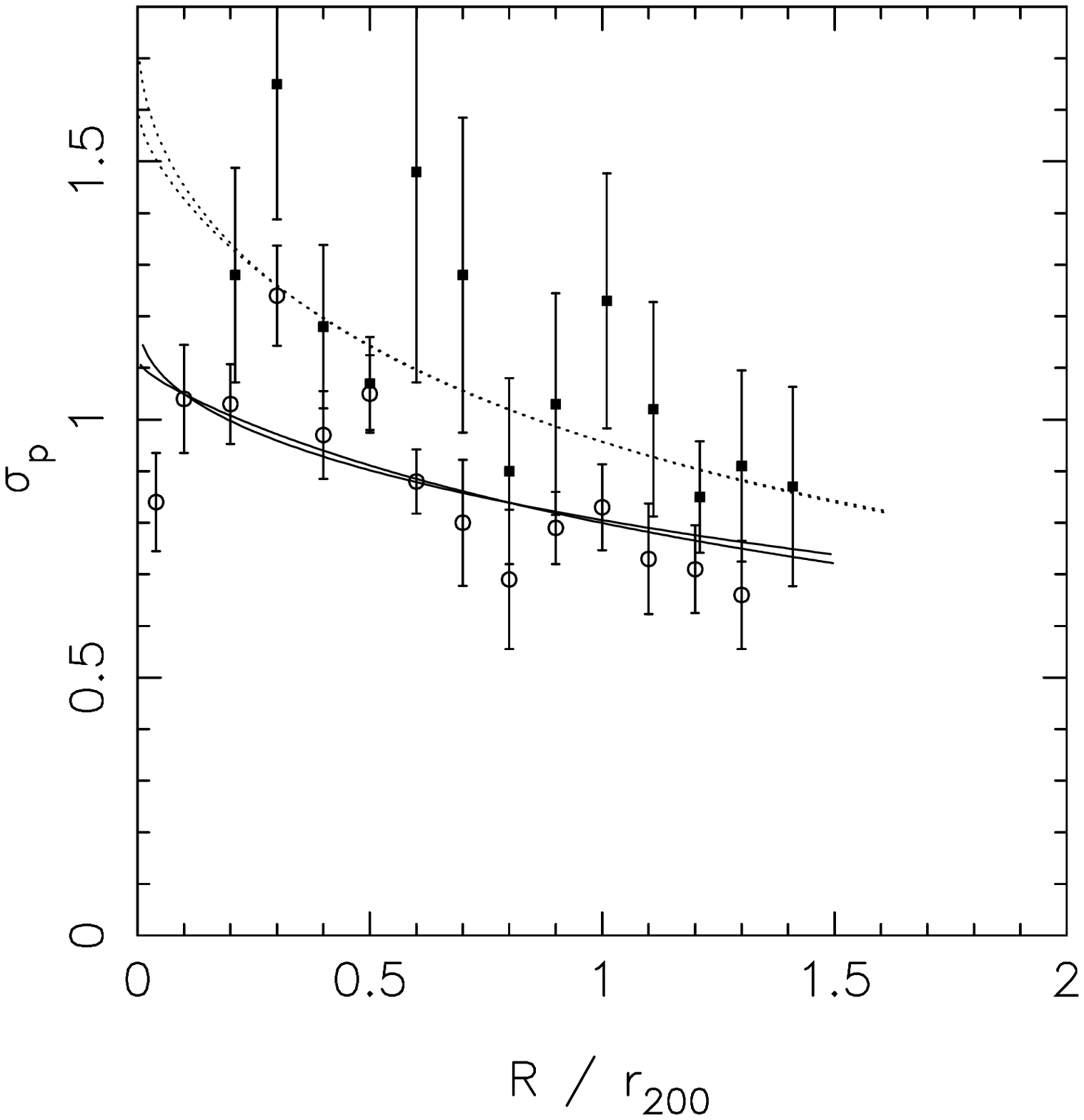}
~~
\epsfysize 4.0truecm
\epsffile{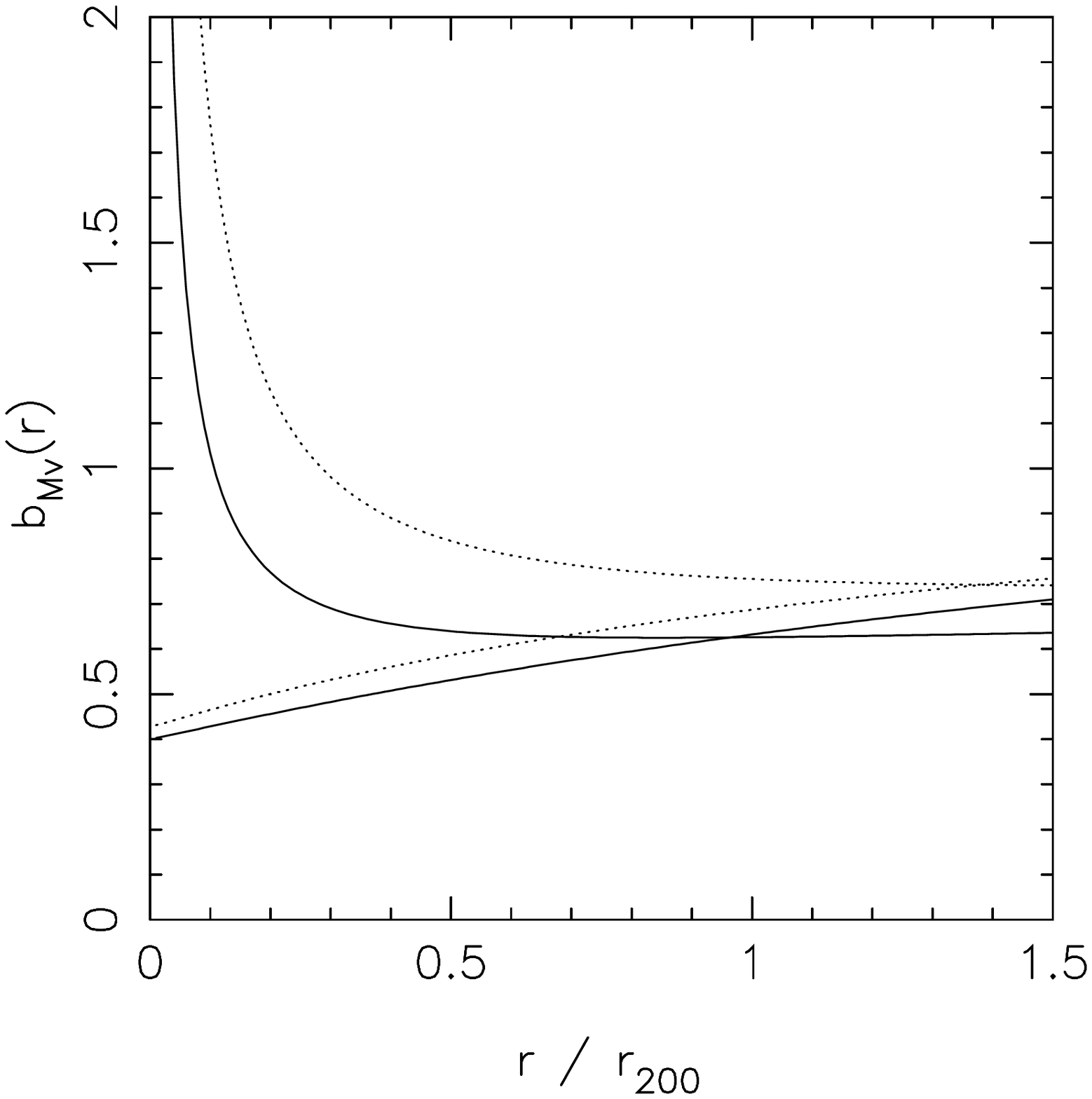}
}
\noindent
Figure 5: The projected number density profiles (left panel) and
velocity dispersion profiles (middle panel) of the blue (solid squares
and dotted lines) and red (circles and solid lines) cluster galaxies.
They yield {\it identical} mass profiles (right panel) near \r200,
where the upper pair of lines are for $\beta=0$ and the lower pair are
for $\beta=0.5$.}
\medskip

Splitting the cluster sample into independent blue and red subsamples
demonstrates that the clusters are effectively in equilibrium and that
the results are robust for drastically different samples. These two
subsamples also illustrate the perils of the virial mass
estimator. The blue galaxies are about twice as extended and have a
20\% higher velocity dispersion than the red galaxies,
Figure~5. Consequently the blue galaxies indicate a virial mass about
three times larger than the red galaxies give. However each of these
subsamples independently gives the same mass profile from the Jeans
equation as found from the full sample, which we take as strong
support for the assumption that both subsamples are effectively in
equilibrium with the cluster potential \cite{br}. It should be noted
that galaxies with red colours are somewhat more concentrated than the
total mass distribution, so the procedure of comparing the mass
profile to the ``red-sequence'' light profile is likely to give a
rising mass-to-light ratio.

The scale radius, $a$, is predicted to be a relatively large 0.20-0.26
of \r200, for simulations normalized to the observed cluster
cosmological density, but otherwise the prediction is relatively insensitive to
cosmological parameters \cite{nfw}. It is a significant success of the
generalized CDM model that the observed scale radius $a=0.13-0.43$
(95\% confidence) is in complete accord with NFW's prediction.

\section{Normalization of the Density Perturbation Spectrum, $\sigma_8$}

The normalization of the density fluctuation spectrum is fundamental
to all predictions of structure formation. The parameter $\sigma_8$ is
the variance of the linear density perturbation spectrum in 8\hmpc\
spheres.  The CNOC cluster sample and observational strategy
\cite{yec} were specifically designed to produce data useful for a
$\sigma_8$ measurement.  The sample's primary advantage is that the
cluster masses are accurately known near the virial radius, which is
essential for a reliable estimate of the linear mass scale from which
the cluster collapsed. We combine our data with similarly selected
clusters from the EMSS \cite{emss2} and ESO Cluster surveys \cite{eso}
to extend the redshift range of the cosmological density estimates.
The cluster cosmological density data are modeled
\cite{ps} to constrain the $\sigma_8-\Omega$ pair. It is not observationally 
straightforward to give accurately the full virialized mass of a
cluster, but it is straightforward to extrapolate the mass inside some
specified aperture, conventionally the Abell radius, 1.5\hmpc. The
measurement of cluster mass inside a fixed aperture is effectively a
mass weighted velocity dispersion or temperature.

\medskip
\hbox{
\epsfysize 4.5truecm
\epsffile{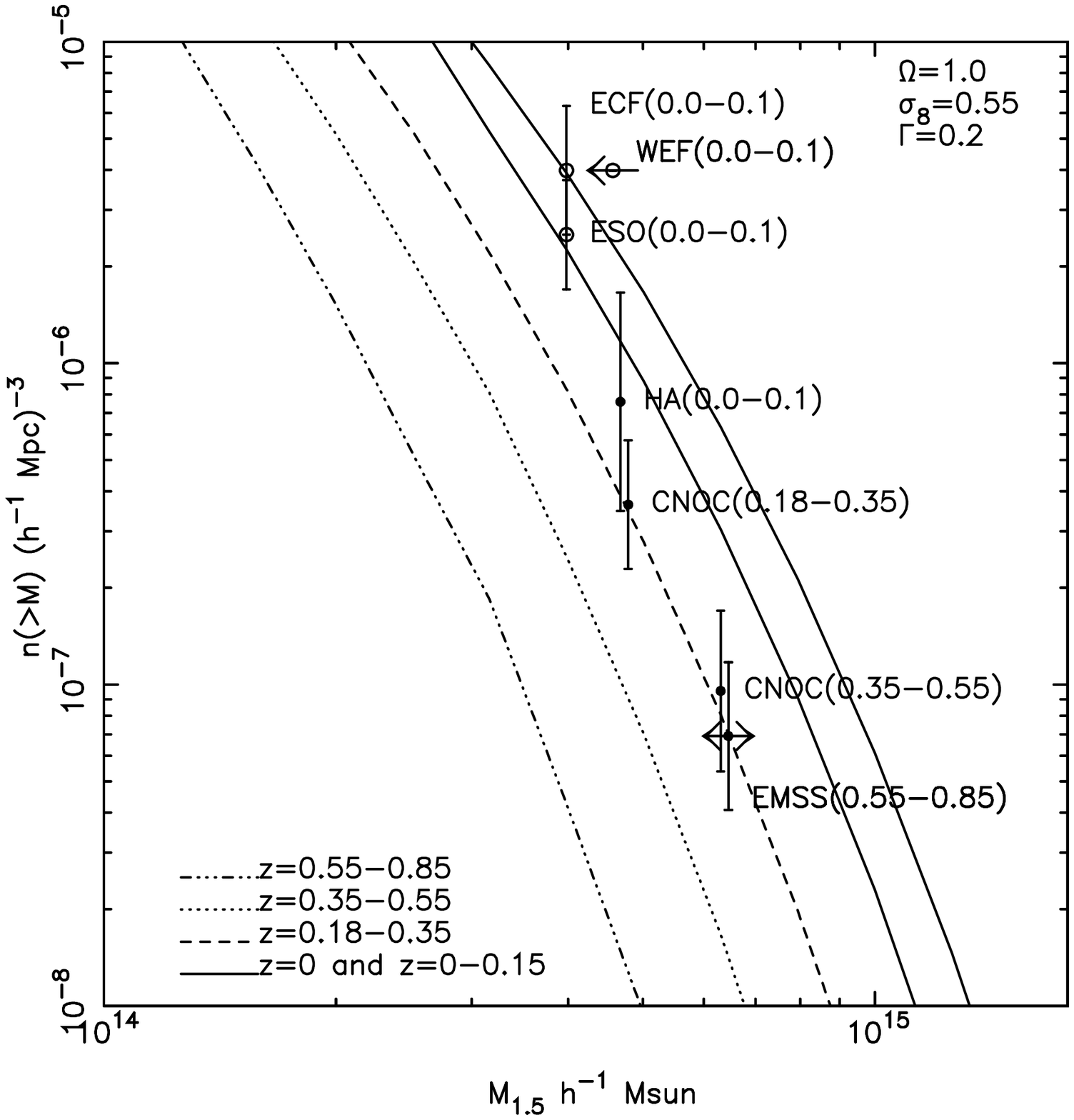}
~\epsfysize 4.5truecm
\epsffile{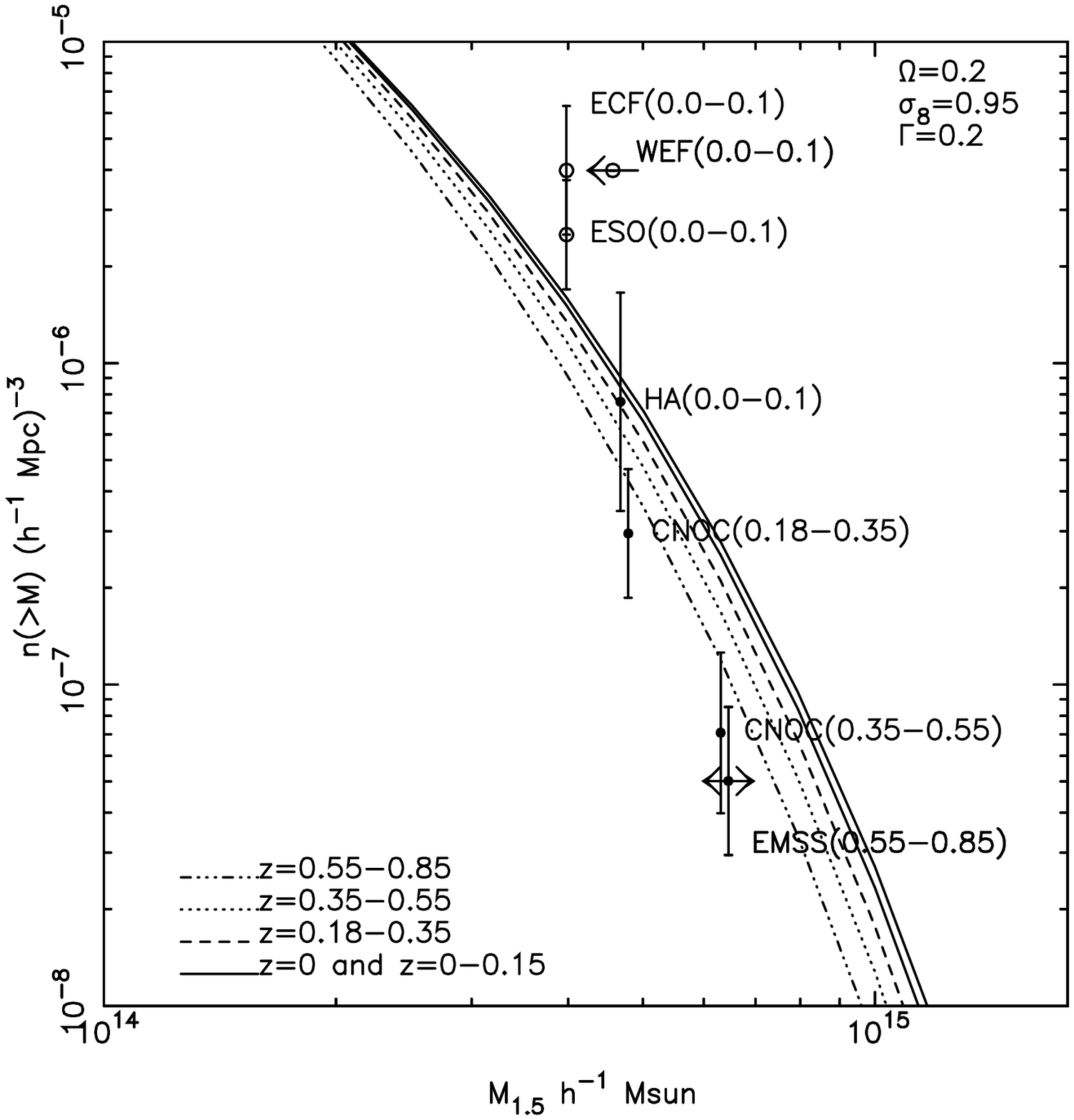}
~\vbox to 4.5truecm{\hsize=3.5truecm
\vfill\noindent
Figure 6: The measured
cosmological density evolution of gal\-axy clusters 
with $\Omega=1$ (left) and $\Omega=0.2$ (right) predictions.
\vfill}}

\medskip
\hbox{
\epsfysize 5.0truecm
\epsffile{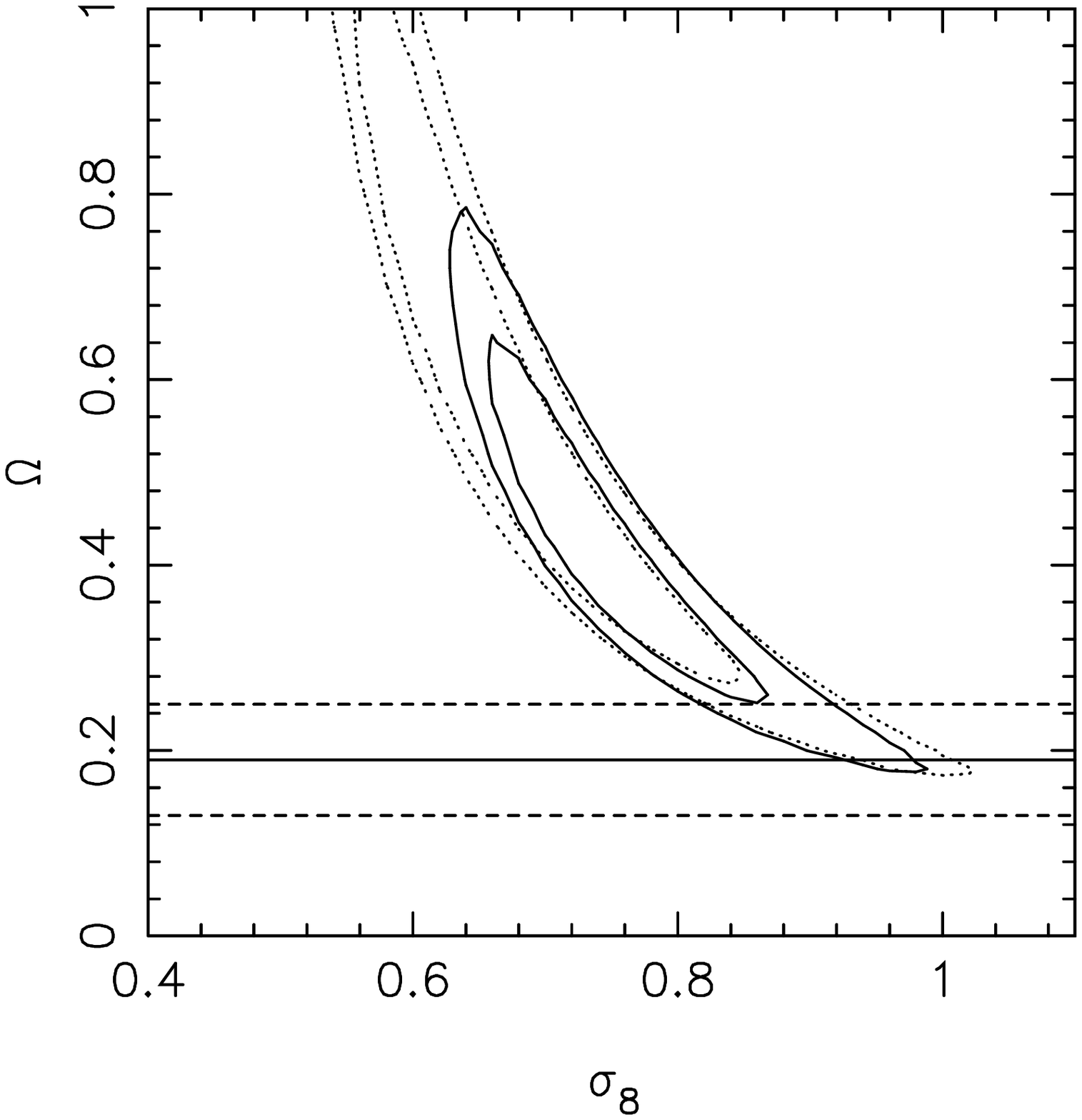}
~\vbox to 5truecm{\hsize=7truecm
\vfill\noindent
Figure 7: A plot of $\chi^2$ for all independent samples (solid lines)
and excluding the high redshift EMSS sample (dotted lines). The
contours are the 90\% and 99\% confidence levels.  The results of the
CNOC analysis, $\Omega=\omzeroc$, with its $1\sigma$ range are
indicated.\vfill  }}

The measured redshift change of the cosmological density of clusters,
having $\sig1\gta 800\kms$ requires some small corrections for the
effects of X-ray selection, which are incorporated in the data of
Figure~6. The analysis is greatly simplified because the $L_x-\sig1$
relation has no detectable redshift evolution, as is also seen in the
X-ray temperatures
\cite{s8,mush}.  The cluster cosmological density data are best described with
$\sigma_8\simeq0.75\pm0.1$ and $\Omega\simeq0.4\pm0.2$ (90\%
confidence), Figure~7. Taking the $\Omega$ from cluster dynamical
analysis as $\Omega=0.2$, we find $\sigma_8=0.95\pm0.1$ (90\%
confidence).  The predicted cluster density evolution in an $\Omega=1$
CDM model exceeds that observed at $z>0.5$ by more than an order of
magnitude, as shown in Figure~6.

\section{Galaxy Evolution}

It is essential for $\Omega$ measurement to understand the
differential evolution of cluster galaxies relative to field galaxies.
The cluster galaxy population is on the average redder than the field,
but the cluster population has an increasing blue fraction with
redshift \cite{bo}. The origin of the blue cluster population has
three basic possibilities. One possibility is cluster galaxies formed
in a different way than field galaxies, with fundamentally different
star formation rates at all epochs. However, hierarchical clustering
leads one to to expect that cluster galaxies originated in the field,
and hence share a similar early star formation history which is
altered in some way upon entry into the cluster. Ultimately star
formation is greatly suppressed in cluster galaxies: however, there
are two routes to this state. Either there was a burst of star
formation upon entry into the cluster which largely exhausted the
galaxy's gas supply, or, the gas was simply swept out of the galaxy
\cite{bs,gg}.

\medskip
\hbox{
\epsfysize 6.0truecm
\epsffile{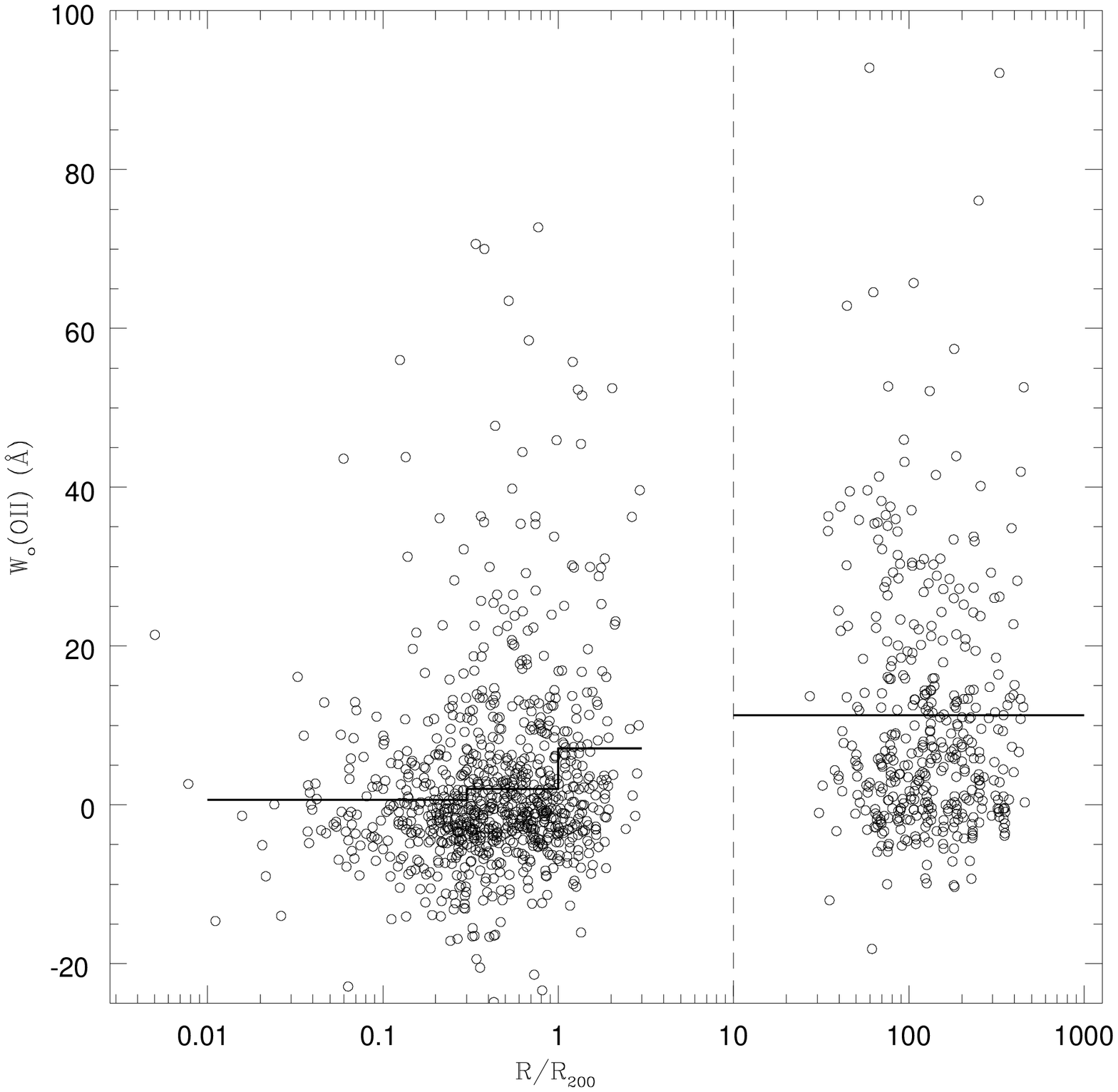}
~\vbox to 5.5truecm{\hsize=6truecm 
\raggedright
\vfill\noindent
Figure 8: The radial dependence of
the equivalent width of [OII] for galaxies with $M_r^k\le -18.5$
(Balogh \et\ 1997 in preparation). The lines give
the mean values in various radial ranges. Nowhere do cluster galaxies show an
excess of star formation over the field. \vfill}}

In Figure~8 we show the equivalent width of the [OII]3727\AA\ line as
a function of radial distance from the cluster center, which is the
projected distance in the inner region and the redshift space distance
for the field galaxies. The sample is limited at a common $M_r^k\le
-18.5$ mag, with an additional cut to eliminate low signal-to-noise
spectra.  The measured equivalent width implies
a mean star formation rate of approximately 
$0.05 h^{-2} \msun$~yr$^{-1}$ in our clusters \cite{kennicutt}.
Over $10^{10}$ years this will add
only a few percent stellar mass to the relatively high luminosity
galaxies in our sample. Of particular note is that there is no
evidence for a significant increase in star formation rate at any
radius. Clusters do contain objects with strong emission lines, mostly
near the cluster ``edge'', but these are rare compared to the
field. We conclude that the difference between cluster and field
galaxy evolution is that star formation is truncated upon the entry of
galaxies into the cluster
\cite{cs,a2390}.

\medskip
\hbox{
\epsfxsize 5.0truecm
\setbox1=\hbox{\epsffile{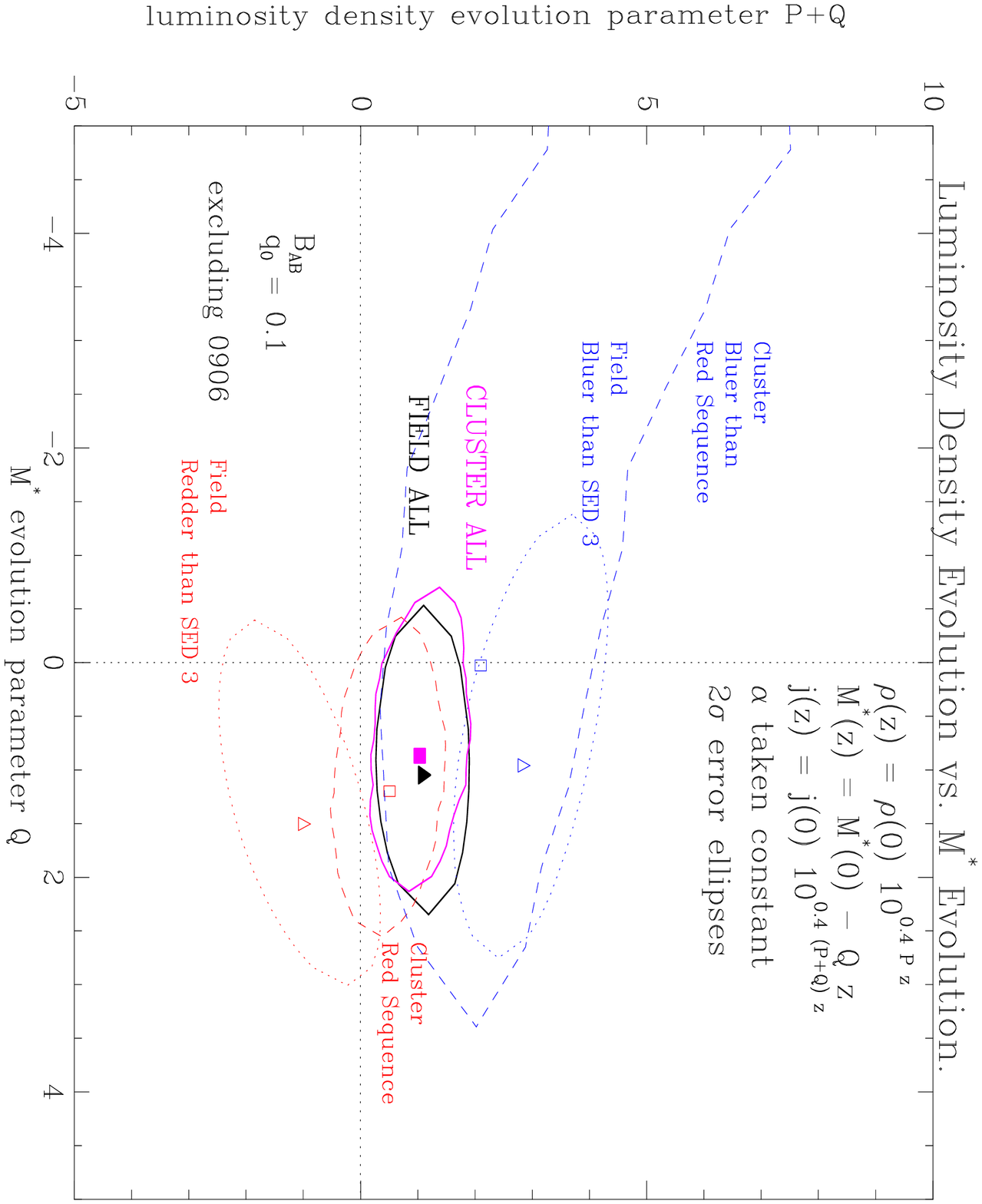}}
\rotl{1}
~\vbox to 5.0truecm{\hsize=6truecm 
\vfill\noindent
Figure 9: The $B_{AB}$-band evolution of $M_\ast$ and $j$ in
magnitudes per unit redshift (Lin \et\ 1997 in preparation). The solid
lines give the 95\% confidence contours for the full samples. The
dotted and dashed lines are for blue and red subsamples.  The cluster
data are, unsurprisingly, consistent with passive evolution. The field
galaxy evolution is nearly identical to the cluster galaxy evolution.\vfill}}

The evolution of the cluster galaxy population should be dominated by
a passively evolving stellar population. The CNOC survey has the
advantage that the masses derived for each cluster can be used to
normalize the luminosity functions of each cluster so that the
redshift evolution of $L/M$ can be measured.  The evolution of the
parameters of Schechter fits to the luminosity functions of galaxies
having $M_B(AB)$ more luminous
than approximately $-17$ mag is shown in Figure~9. The $Q$ parameter
measures the evolution of $M_\ast$ with redshift (the faint end slope
is held fixed at $\alpha=-0.83$, the best fit value) as
$M_\ast=M_\ast(0)-Qz$. The evolution of the luminosity density,
$j(z)$, is nearly statistically orthogonal to $M_\ast$. We model
$j(z)$ as an evolution in magnitudes, $j(z)=j(0)10^{0.4(P+Q)z}$.
These luminosity function and evolution parameters are found using a
parametric maximum likelihood technique \cite{sty,saunders} applied to
1838 field galaxies and 1018 cluster galaxies above the sample limits.

We find that to a good approximation the density evolution parameters
$P\simeq0.17$ (consistent with zero) indicating that there is little
density evolution of cluster galaxies, as we expect for these high
velocity dispersion clusters where merging and star formation are
insignificant. We find $Q\simeq 0.87$, which n accord with predictions
of purely passive evolution of a predominantly old stellar population
of current at $\sim$15 Gyr today. In the field ($\alpha=-0.93$), we
find $Q\simeq1.05$ and $P\simeq0.04$, nearly identical to the cluster
values. Nearly identical values are inferred from this approach to the
analysis of the CFRS data
\cite{cfrslf,lin,cfrsj}. The mean field star formation rate that we infer from
[OII] over the $0.2 \le z \le 0.6$ interval is only about $0.3
\msun$~yr$^{-1}$ for galaxies more luminous than $0.2L_\ast$ with
a weak redshift dependence. This low mean star formation rate would
add about 5-10\% more stellar mass to an $M_\ast$ galaxy (close to the
sample mean and median) over this redshift interval, consistent with a
mild field luminosity evolution and a modest difference between
cluster and field.  These results are supported by direct,
cosmological model independent, measurements of the surface brightness
of the galaxy spheroids and disks \cite{schade_e,schade_d}.

We conclude that field galaxies evolve roughly in parallel to cluster
galaxies, although cluster galaxies have lower luminosity $M_\ast$, by
$0.3\pm0.1$ mag in B, or using different procedures, $0.11\pm0.05$ mag
in $r$. Simple stellar population modeling for truncated star
formation shows that this is consistent with the mean colour
difference of about 0.2 mag. The fading of cluster galaxies relative
to the field is an essential, but relatively small, correction in the
$\Omega$ estimate.  We interpret the Butcher-Oemler effect as likely
being the result of blue, star forming, field galaxies falling into
clusters at a rate that increases with redshift, although this
remains to be quantified and tested.

\section{A Neo-Classical $\Lambda$ Estimator}

Cluster data allow a measurement of the geometry of the universe,
using the following procedure.  The mass density of the universe,
$\rho_0$, is a conserved quantity which is estimated with the product
of $M/L(z)$ and $j(z)$.  The $M/L(z)$ is estimated from the clusters,
with a relatively small correction, 10-30\%, to allow for differential
evolution with respect to the field. The luminosity density, $j(z)$,
involves the volume element, which is strongly $q_0$ dependent. If
these quantities are calculated using the values of $\Omega_0$ and
$\Omega_\Lambda$ that the real universe has, then $\rho(z)=M/L(z)
\times j(z)$ will be constant at $\rho_0$. However, if the ``wrong''
values are assumed, then the calculated $\rho(z)$ will be dependent on
redshift. For instance, if we assume that $\Omega_0=1$, but in fact
$\Omega_0=0.2$, $\Omega_\Lambda=0.8$, then $\rho(z)$ will rise 76\%
from z=0.2 to z=0.8, which is easily detectable in samples of the CNOC
size but having a larger redshift range. Note that this effect exceeds
the entire differential evolution between field and cluster by a
factor of two, and that the redshift variation in the differential
evolution is even smaller. Although we will correct for galaxy
evolution effects, they are small compared to the geometry changes we
are trying to measure.

\section{Discussion and Future Directions}

The main goal of the CNOC survey is to use clusters of galaxies to
derive a value of $\Omega_0$ with a well determined error budget.  The
major innovation of our analysis is that it is completely
self-contained, with the key assumptions being testable, and that the
error estimates are derived from the data themselves.  The dominant
source of error is random cluster-to-cluster variations, rather than
the internal error from individual clusters.  The global mass-to-light
ratio (in our photometric system, at mean redshift of $z=0.31$, with
k-corrections, but without evolution corrections) of our sample
clusters of galaxies is constant within our typical errors of 25\% at
a value of $\mlobs h\msun/\lsun$. Over the same redshift range we
measure the closure value, $\rho_c/j$, to be $\mlclose h\msun/\lsun$
\cite{global}.  After allowing for the $\ladj$ mag lower luminosities of
cluster galaxies and reducing $M_v$ by $\mvadj$, we find that
$\Omega_0=\omzeroc$. In an independent luminosity function analysis,
the evolution corrected $B_{AB}$ value of $M_v/L$ leads to
$\Omega_0\simeq0.23$ with a similar error budget.

There are no variations of the radial mass-to-light profile within the
clusters, nor is there any variation with redshift. This requires any
additional form of dark matter to be sufficiently hot that it does not
fall into clusters, but perhaps participates in large scale
flows. This is a very narrow parameter range and produces trouble for
the evolution of the density perturbation spectrum. The most likely
form of additional mass-energy, if it exists, is in the form of
$\Lambda$, which additional cluster plus field observations in a
neo-classical volume-redshift test can easily detect.

Two aspects of the cluster mass evolution are particularly satisfying.
The derived $\Omega_0$ accurately predicts the observed evolution of
cluster cosmological density. With our value of $\Omega_0$ we find
that $\sigma_8=0.95\pm0.1$.  Moreover the same $\Omega$ in a cool,
dark matter dominated universe, predicts remarkably accurately the
``morphology'' of the clusters, at least as encapsulated in the NFW
results for the mean mass profile.  For this low density cosmology the
measured rate of both structural evolution, Figure~10 \cite{ccsh}, and
galaxy evolution, Figure~9, over our redshift range is consistent with
a formation ``freeze out'' at some $z>1$, and possibly much higher.

There are considerable opportunities to refine and extend these
results in a number of different directions. For the current sample
these include improved coverage at 1-3\r200\ to constrain the
luminosity profile at large radius, which would lead to a tighter
comparison with the core radius predictions and the slope at large
radius. The same data would advance the empirical study of how field
galaxies are altered upon infall into the cluster. The turnaround
radius is at about 5\r200. If the sample extended to projected radii
well beyond turnaround, then there would be a clear measurement of the
infall into the cluster, which is a ``bulk flow'', hence measures the
$\beta=\Omega^{0.6}/b$ parameter. Such a measurement would directly
compare the cluster $\Omega$ estimate with a field $\Omega^{0.6}/b$
measurement to resolve the large scatter present in the current
estimates from flows.  At these large radii many of the galaxies would
be distant field galaxies, but their immense value is to provide a
good estimate of the unperturbed background density, which is an
essential ingredient in the measurement. Including the four high
redshift EMSS clusters would allow an $\Omega_\Lambda$ measurement
with a precision of about $\Delta\Omega_\Lambda\simeq0.25$ to be
made. With an efficient multi-object spectrograph on a 4 meter
telescope, it takes 1-2 nights to observe a cluster at $z\simeq0.8$.

\medskip
\hbox{
\epsfysize 6.0truecm
\epsffile{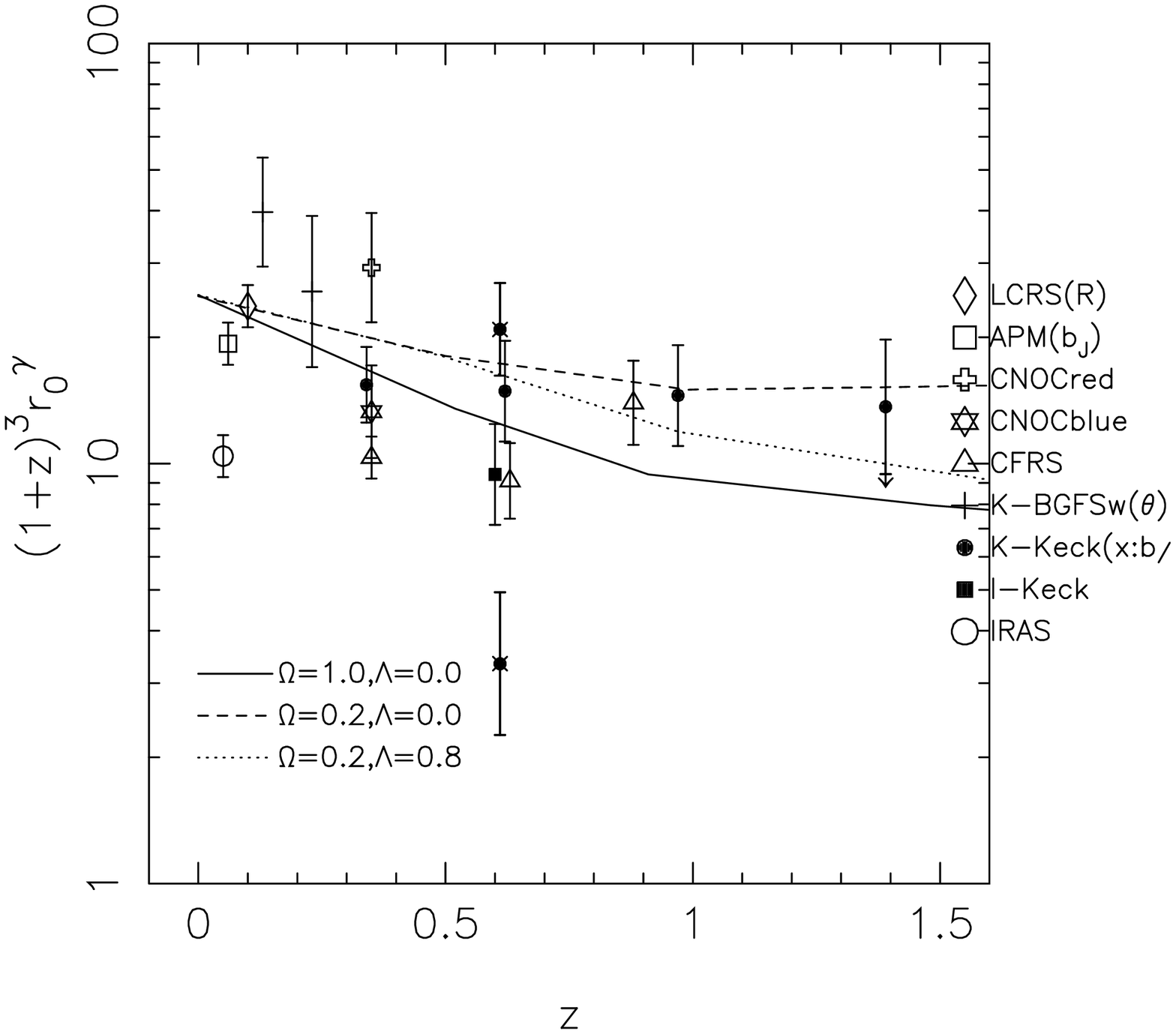}
~\vbox to 6.0truecm{\hsize=5.5truecm 
\noindent\raggedright\vfill
Figure 10: The evolution of the physical density of clustering with
redshift from various samples. Red galaxies are more correlated than
blue ones, notably the K-Keck \cite{ccsh} and CNOC2 field (Shepherd
\et\ 1997, in preparation) samples. There is no compelling evidence for
any change of physical density of clustered galaxies with redshift.\vfill}}

More ambitiously one can imagine in the near future assembling much
larger cluster samples to improve the precision of these cosmological
parameter estimates to the few percent level that Cosmic Microwave
Background experiments hope to attain.  For instance, a sample of
200-300 clusters, with the accompanying field, spread more or less
uniformly over the $0<z<1$ range would allow $\Omega_M$ to be measured
to a precision of about 2-3\% and $\Omega_\Lambda$ to be measured to
about 5\%. The $\sigma_8$ parameter could be measured to a precision
of better than 1\%, which would become a very strong constraint on the
overall spectrum of fluctuations. The galaxy evolution parameters
would be measured at about 5\% precision and would challenge stellar
population models, and might even begin to rival globular clusters as
tools to measure the age of the universe. However, this would also take
some advances in modeling precision.  The realistic errors are likely
to be about twice as large, since new residual systematic errors will
be uncovered and removed as the dataset grows.  At redshifts beyond
about $z=0.5$ the first problem is to find clusters in some systematic
manner. This is mainly a matter of systematic sky surveys of various
sorts. It should be noted that, in the $0.2\le z \le 0.55$ range of the
EMSS/CNOC survey, we selected only 15 clusters that are sufficiently
rich that they are easy to study from 584 square degrees of sky. The
implication is that these large, rich, easily-studied clusters are
rather scarce on the sky. Nevertheless, once found, clusters of this
sort will be exceptionally powerful as direct probes of both
cosmological processes and cosmological parameters. Such programs will
be feasible with a number of the new generation of telescopes.

\end{document}